# ESASCF: Expertise Extraction, Generalization and Reply Framework for Optimized Automation of Network Security Compliance

**MOHAMED CHAHINE GHANEM[1 and 2], (Member, IEEE), THOMAS M. CHEN[3], (SENIOR MEMBER, IEEE), MOHAMED AMINE FERRAG[4], (Senior Member, IEEE) and MOHYI E. KETTOUCHE[1]**

[1]Cyber Security Research Centre. London Metropolitan University. London, UK
[2]Department of Computer Sciences, University of Liverpool. Liverpool, UK
[3]Department of Engineering. City, University of London. London, UK
[4]Technology Innovation Institute, Abu Dhabi, UAE

Corresponding author: Dr Mohamed Chahine Ghanem (e-mail: ghanemm@staff.londonmet.ac.uk).

**ABSTRACT** Organizations constantly exposed to cyber threats are compelled to comply with cyber security standards and policies for protecting their digital assets. Vulnerability assessment (VA) and penetration testing (PT) are widely adopted methods for security compliance (SC) to identify security gaps and anticipate security breaches. However, these methods for security compliance tend to be highly repetitive and resource-intensive. In this paper, we propose a novel method to tackle the ever-growing problem of efficiency in network security auditing by designing and developing an Expert-System Automated Security Compliance Framework (ESASCF). ESASCF enables industrial and open-source VA and PT tools to extract, process, store and re-use the expertise in similar scenarios or during periodic re-testing. ESASCF was tested on different size networks and proved efficient in terms of time efficiency and testing effectiveness. ESASCF takes over autonomously the SC in re-testing and offloading the human expert by automating repeated segments SC and thus enabling experts to prioritize important tasks in ad-hoc compliance tests. The obtained results show a performance improvement by cutting the time required for an expert to 50% in the context of typical corporate networks' first security compliance and 20% in re-testing. In addition, the framework allows a long-term impact illustrated in the knowledge extraction, generalization, and re-utilization, which enables better SC confidence independent of the human expert skills, coverage, and wrong decisions resulting in false negatives.

## I. INTRODUCTION

In the digital age, our daily routines are increasingly reliant on the security and resilience of various Internet-connected devices and computer systems. However, the convenience of ubiquitous computing comes at a price as computer networks continue to grow in size, complexity and interconnection to perform a wide range of tasks for the benefit of users and organizations. Along with this expanding connectivity, cyber threats are becoming more frequent, complex and sophisticated, providing cybercriminals with more opportunities to launch malicious attacks attempting to gain access to sensitive data for their benefit [1], [9]. Being flexible comes at a huge cost, as cybersecurity professionals, experts, and researchers have found that cyber threats are becoming more frequent, complex, and sophisticated as the general rule of the attack surface evolves. Protecting complex networks and critical assets from cyber threats has forced network security professionals to implement more and more security layers and policies [6]. The defence-in-depth approach is complex and results in the addition of multiple layers of security that are often vulnerable to a high-level attacker because they contain vulnerabilities due to human error, misconfigurations, and system weaknesses. Therefore, the primary concern of cybersecurity communities is to ensure that the security measures applied are effective [26].

Several approaches have been proposed and adopted over time. Nevertheless, using the offensive approach has proven to be the best and most reliable method, and has received the most positive reception from cybersecurity professionals [40]. At its core, cybersecurity compliance is a well-established security auditing method that aims to ensure adherence to standards, regulatory requirements, and laws [34]. Since the introduction of GDPR and related legislations across the world, organizations are legally required to achieve compliance by establishing risk-based controls that protect the confidentiality, integrity, and availability (CIA) of their digital assets (computers, networks, web applications, servers, etc.) by trying to identify vulnerabilities and measuring the associated risk [1].

In this paper, we are concerned with making security compliance and penetration testing more efficient by enabling industrial tools and systems to observe, capture and replay human expertise in future cases relying on a novel representation of the practice and the use of knowledge-based and rule-based expert systems.

### A. BACKGROUND ON SECURITY COMPLIANCE

Security compliance constitutes a central and mandatory component of the cyber-security audit and embeds all standard auditing and testing tasks starting from information gathering, analysis, planning, and testing the appropriate attacks targeting the identified vulnerabilities [36]. Such assessments are considered the most effective method to identify exactly how effective the existing security controls are against a skilled adversary and validate the efficacy of defensive mechanisms, as well as end-user adherence to security policies [21].

ISO/IEC 27001 is a neutral and worldwide approved standard for information security management systems (ISMS), along with PCI-DSS (Payment Card Industry Data Security Standard) in the financial sector and HIPAA (Health Insurance Portability and Accountability Act) in the healthcare sector, they constitute a cornerstone of security compliance standardization. Security compliance is formalized through these three industry standards, namely ISO-27001, PCI-DSS, and HIPAA, and designed to be a comprehensive and multi-phase practice carried out by experts which usually involves the use of versatile tools, systems, and frameworks to accomplish different tasks [14].

For instance, the information gathering phase typically involves utilizing tools such as traffic monitoring, port scanning, and OS fingerprinting to gather relevant information that can be used to dress the target system defences and determine if it contains a vulnerability that can be exploited [20]. On the other hand, the exploitation phase (if required) employs a set of frameworks, add-on modules, and scripts in order to customize and execute the selected exploits which can vary from pieces of code to data payload with the ultimate aim of taking advantage of the discovered vulnerability and causing unintended behaviour in the system or compromising the target leading to gain additional privilege access [32] and [31].

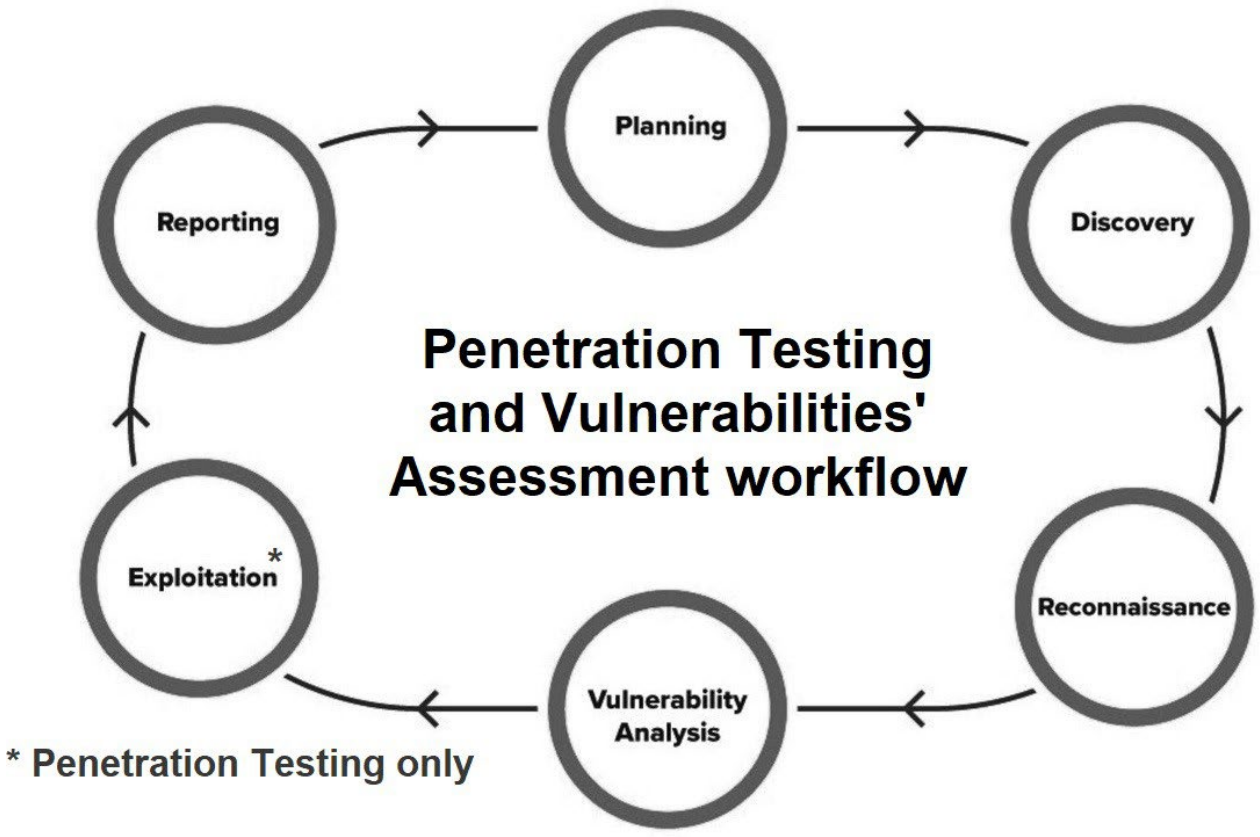

FIGURE 1: Penetration testing and vulnerability assessment are standard methods for assessing network defence and achieving security compliance by following sequential and interactive multi-phase procedures starting by gathering information and ending by reporting the obtained results.

In addition, once an exploit execution is successful, post-exploitation tools and frameworks are heavily utilized in order to maintain the breach and work toward further penetration [10]. Finally, SC also involves versatile testing scenarios and contexts with tested assets that differ immensely. In each case, the same general phases are followed but executed tasks differ significantly [10], [21]. VA and PT are methodological approaches which involve an active extraction, analysis, and exploitation of the assessed assets and their potential vulnerabilities [39]. Being the industry's standard security compliance method, PT and subsequently VA rely on a set of classic tools that automate repetitive and complex tasks [29]. The PT tests are often initiated and carried out from the position of a potential attacker and involve active exploitation of security vulnerabilities. Real-time exploration and decision-making as the practice evolves are the key [33].

The human expert's knowledge, decision-making, and reasoning are a cornerstone of the PT and VA [11]. Currently, PT and VA tools and systems are developed to make the practice efficient and allow regular and systematic testing without a prohibitive amount of human labour along with reducing the precious consumed time and network downtime [20]. Additionally, they are designed to offload human experts from heavy tasks and helping him/her to focus on more special and complex situations such as unusual vulnerabilities or combined non-obvious combinations making improper configurations, and risky end-user behaviours) which require particular attention in order to produce the best results [23].

Additionally, the wide variety of assets and vectors such as servers, endpoints, web applications, wireless networks, network devices, mobile devices and other potential points of exposure are playing against the pen-tester breaking through the network firewall and evolving beyond by pivoting across networks machines, systems and applications and attempting to find a new path of attack or revealing how chains of exploitable vulnerabilities to progress further within the target network critical systems and data [33]. Figure 2 illustrates the versatility of security compliance practice.

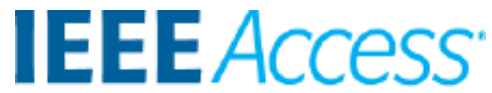

### B. RESEARCH MOTIVATION

This research examines the practical issues that professionals in the offensive cyber security field frequently encounter notably with the pressing demand for PT, making it a compulsory part of cyber-security audits, aligning with various global norms and regulations. The study aims to provide a scientific remedy by examining current automated procedures, selecting the most suitable knowledge extraction methods to integrate it with an expert system and offering an efficient, flexible and universal PT framework. This framework would conduct an optimized penetration test in the context of a network, whilst being self-sufficient and capable of self-learning [26].

The VA and PT practices have significantly evolved to keep pace with cyber advisories, and this led to the appearance of dozens of commercial and professional systems and frameworks which all aim to offer automation of the different activities, tasks and sub-tasks [7], [10]. Nonetheless, the existing automation remains either local (specific to one activity such as the vulnerability scanning) or not optimized (covering blindly all cases including irrelevant ones). These reasons make current VA and PT systems such as Metasploit and Nessus being used as tools fully controlled by the expert and only executing tasks launched by the human according to his/her decisions which often lacks prioritization and optimization. The expert uses output to analyze, plan and request the execution of the required tasks and those systems only execute the expert instructions [8].

Furthermore, the repetitive nature of security compliance practice is becoming problematic, especially during periodic or ad-hoc compliance where most of the workload remains unchanged, and this problem worsens in large IT assets [16]. All the reasons enumerated in this section triggered this research and the expert system choice is backed by the lack of knowledge extraction, re-usability and improvement as is the case during manual security compliance which is the main reason behind expert VA and PT poor efficiency [2].

### C. RESEARCH CHALLENGES

All organizations across the world are witnessing an increase in terms of connectivity and online resources making a higher number of machines exposed online and thus a larger attack surface [11], [26]. Attacks can range in scale from massive state attacks to simple attacks on individuals and SMEs in the hopes of gaining credentials or financial details [3], [23]. In addition, other issues arise with the use of such automated systems in combination with issues raised on the manual approach notably:

1) The high cost of regular and ad-hoc security audits in terms of human resources and cost, consumed time and the impact on the IT assets' performances and systems downtime during working hours.
2) The high volume in terms of data produced by comprehensive non-targeted testing is often wasted and unexploited properly.
3) The nature of the PT environment is where the high threats' emergence and fast-changing rate along with assets' continuous security protection evolution.
4) The evolving attacks' complexity with more evasive threats launched in which hackers adopt complex and indirect attack routes, techniques, and technologies, results in unlikely paths being used to squeeze through the security layers which is difficult to imitate during PT and VA.
5) The a huge amount of repeatability as most of the performed activities and tasks are repeated with hardly any change. This represents a significant part of testers' time, often repeating does not require PT human expert decision-making or manual intervention which results in decreasing the performances.
6) The common high degree of obfuscation in large infrastructures notably in the corporate and financial sectors where organizations tend to use in-house developed security systems makes the coverage of the whole assets challenging.

## II. METHODOLOGY

This section provides an outline of the research methodology followed and the chosen approaches towards an ES-led security compliance framework. This research started by reviewing the state of the art in the domain of VA and PT automation and optimization, identifying key elements of the current practice requiring improvements [21], [26]. This survey and critical evaluation of existing methods led us to consider the suitability of many AI techniques to settle down on a rule-based expert system and then proceed with designing, developing, testing and evaluating the proposed ESASCF. In summary, the proposed methodology is expected to address scientifically the real-world problem of efficiency and effectiveness related to the current VA and PT automation. The research methodology's five steps are summarized as follows:

- Grasping the VA and PT domains and components and understanding the interaction between the different entities and the human expert.
- Reviewing the current state of the art of the current methods of VA and PT automation at different phases of the practice such as information gathering, discovery, vulnerabilities assessment and exploiting to fully digest and analyze the functioning mechanisms of each and the reason why they fail to meet the PT expectation in term of efficiency and accuracy.

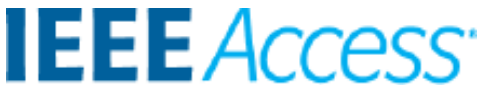

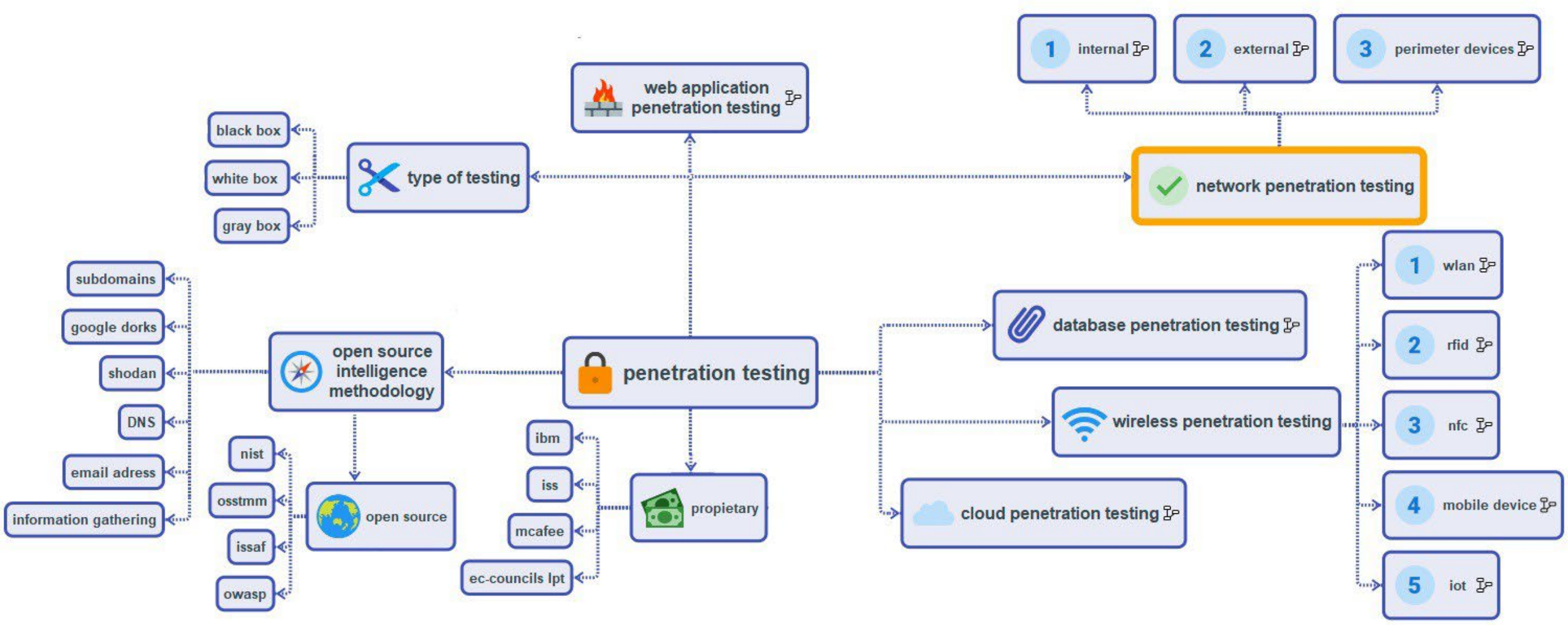

FIGURE 2: The versatility in penetration testing and vulnerability assessment in terms of tasks, methods and domains of practice.

- Studying the cyber security auditor and experts' (eg. Certified Ethical Hackers) methods, operations and approaches when performing security compliance tests. This includes a detailed understanding of activities, tasks and sub-tasks that experts perform from the initial reconnaissance and data gathering to the exploiting and post-exploitation tasks.
- Investigating the suitability of rule-based reasoning and how the expert system can reduce or even replace human intervention in the sequential decision process in VA and PT and which approach is more suitable and likely to produce results.
- Producing an initial expert system using CLIPS which capture, process, generalize and reuse expertise from human-led network PT and VA activities. The developed ES is then integrated as a separate module within ESASCF.
- Testing the proposed solution and evaluating its contribution in terms of efficiency and accuracy in real-world large security compliance cases and subsequently introducing the appropriate changes in due course.

This adopted methodology aims to achieve the research's final output which is a novel ES-led security compliance framework ESASCF that will offload the human expert in performing security compliance and covering the entire spectrum of activities, tasks and sub-tasks [26].

## III. EXPERT SYSTEM FOR SECURITY COMPLIANCE

### A. EXPERT SYSTEMS OVERVIEW

An expert system is a rule-based decision tree program that utilizes AI technologies to simulate the judgment and behaviour of a human or an organization that has expertise and experience in a particular field [4]. Expert systems are usually intended to complement and not completely replace human experts [28] and [11]. Expert systems intended to model human expertise or knowledge by learning either by receiving (implementation) or capturing the expertise or knowledge directly from human experts while being aware of the environmental parameters under which these later have been taken [2].

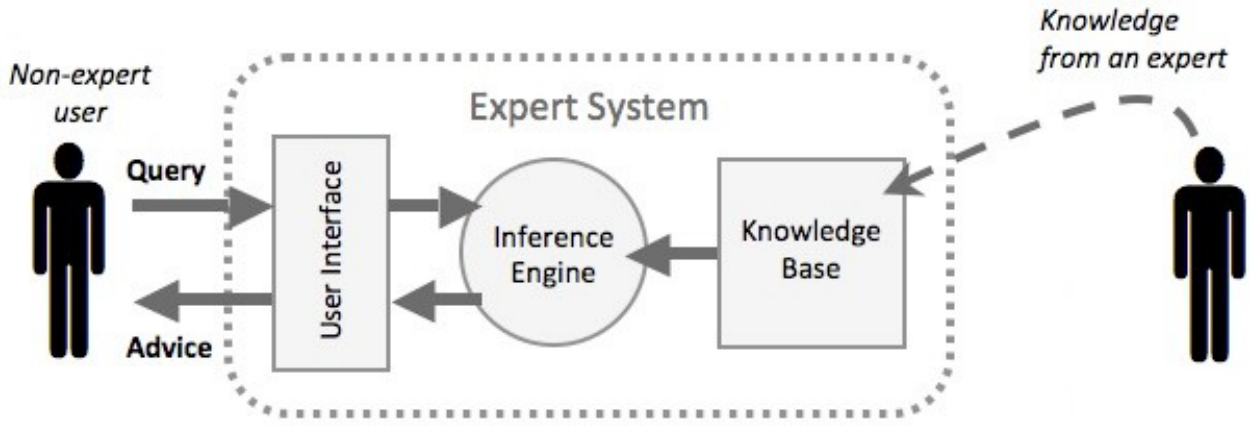

FIGURE 3: Expert system functional diagram in the context of human assistance.

In practice, this is done typically in three different ways:

1) Rules: these are mainly intended for capturing and modelling human expert decision-making in the form of a state-action format which reflects knowledge representation based on experience.
2) Functions: defined and generic functions which are primarily intended for procedural knowledge.
3) Object-oriented: this is programming oriented mainly intended for procedural knowledge with accepted features including classes, message handlers, abstraction, encapsulation, and inheritance.

The C Language Integrated Production System (shortly annotated as CLIPS) is an expert system-building tool, a simple and complete environment for the development and implementation of rule-based expert systems [18]. CLIPS is particularly efficient and is designed to provide a low-cost option for deploying expert system applications across resource-constraint hardware platforms. Following its first release, CLIPS has undergone several upgrades and improvements to become one of the most attractive rule-based expert systems in applied research works. CLIPS's main strength is its ability to facilitate software development to model human knowledge or expertise [27].

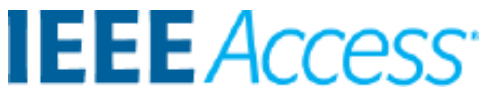

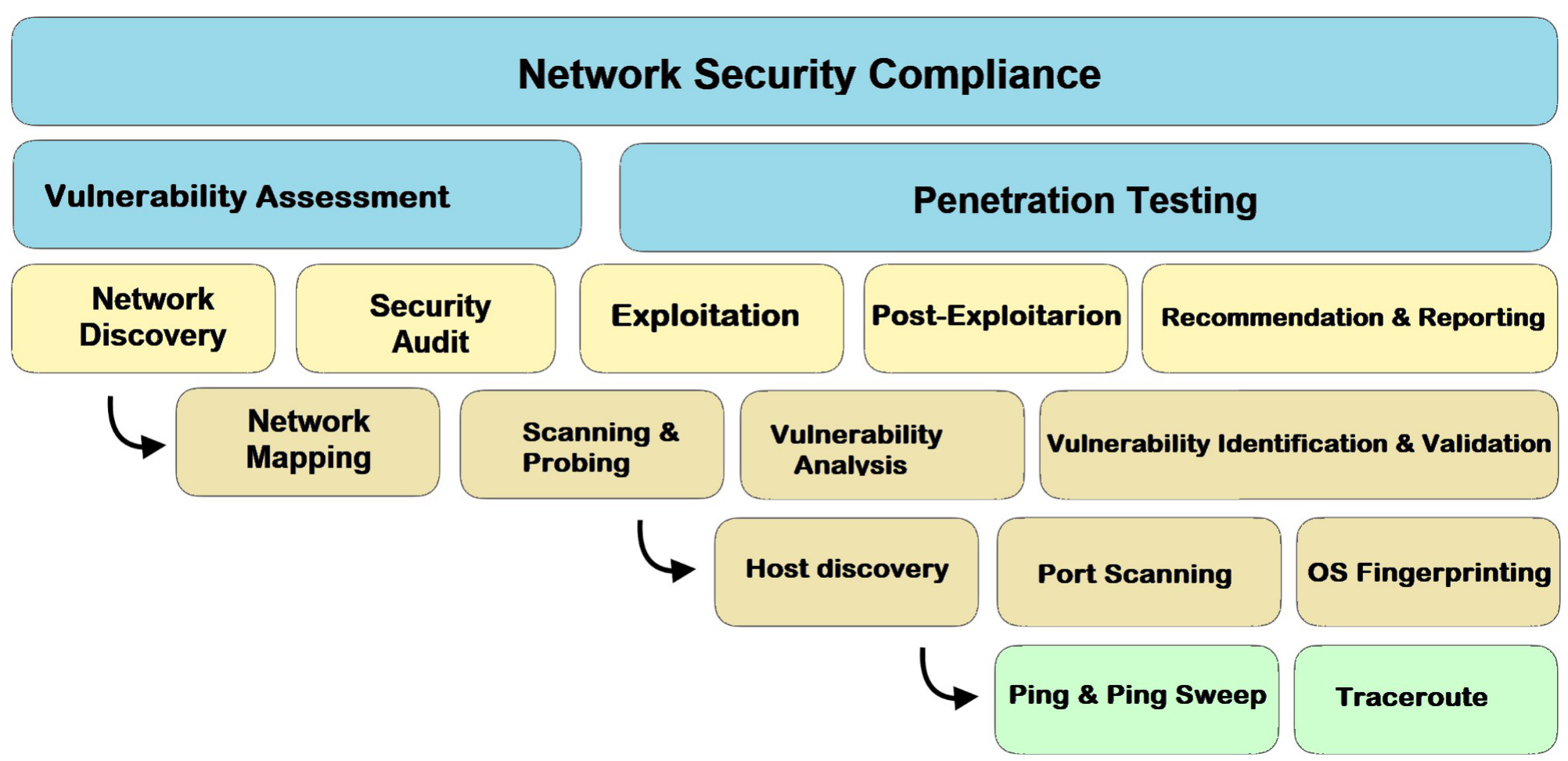

FIGURE 4: Proposed representation of Cyber Security Compliance in the form of activities, tasks and sub-tasks

### B. SECURITY COMPLIANCE EXPERTISE MODELING AND REPRESENTATION

In this subsection, we will detail the method used in our research to model SC activities, tasks and sub-tasks as processes. We will also detail the representation of this expertise in the form of rule-based ES inspired by a deep understanding of the human technical expertise and knowledge role in the VA and PT practice [30]. This enabled us to implement these activities, tasks, and sub-tasks in a CLIPS expert system. The activities in VA and PT are divided into a sequence of tasks in order to methodically and comprehensively identify existing vulnerabilities and perform a set of tasks to assess and test if the target is vulnerable or could be compromised by running exploits against identified vulnerabilities [41].

In our quest to design the CLIPS expert system, we followed a rigorous examination of the security compliance activities, tasks, and sub-tasks. In fact, at this stage, we attempted to grasp the domain fully. We noticed that VA and PT experts adopt a multi-phase operating mode which includes reconnaissance, vulnerability scanning, identification, validation, and optionally exploitation for all computers, equipment, networking, and security devices constituting the assessed network [29]. As a result, we concatenated previous research output and elaborated a novel universal workflow that accounts for and represents all activities, tasks, and subtasks in network security compliance as shown in Figure 5.

We introduce here a novel algorithm that constitutes the main component of ESASCF and covers the expertise identification, extraction and validation based on predefined criteria. In practice, this algorithm process is virtually separated into two tasks which consist of extracting the expertise in the form of attack vectors and then evaluating this expertise compared with past similar expertise and only validating if it exceeds the past expertise in terms of the likelihood of being the optimal decision flow as explained in figure 10.

We define the following notions:

- S is the network state space including topology, machine configuration, and running services details.
- A is the possible actionable tasks and sub-tasks that the SC expert can perform.
- E and V are respectively the list of possible exploits and vulnerabilities that apply to the network context imported and processed from the CVE database.
- C is the possible state of compromised machines within the network.

### C. RULE-BASED EXPERT SYSTEM FOR SECURITY COMPLIANCE

We detail here the method adopted into the definition of expertise from PT and VA perspectives. The proposed rule-based expert system takes knowledge from a human Certified Expert Hacker and converts it into a set of hard-coded rules to be applied in future tests which will ultimately result in fully autonomous PT systems that rely on a well-defined expert system in emulating the decision-making ability of a human expert. The proposed ES will be developed in a modular way to enable future integration with previously developed modules to form a proof-of-concept ES-led Automated Security Compliance Framework (ESASCF). In order to put the ES into practice, the definition of SC expertise definition is the cornerstone of the process. We opted for the most realistic method of defining expertise mimicking the human experts and respecting the PT and VA workflow as illustrated in Figure 6.

The proposed rule-based expert system is written in CLIPS which is a data-driven program where the facts, and objects if

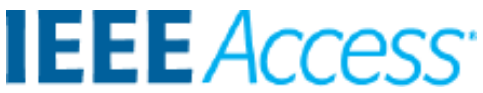

```
Input: S, A, E, V, C, R #Max vector length.
Output: G # Actions' Vector for state s, L# Likelihood L for the V actions' vector at state s to be optimum.

for s ∈ S do
                        /* Optimal Sub Vector extraction */
    E_s ← EXTRACT (V_s, (MAX_Prob (C_s, E_s)), A_s-1)

            /* Extraction of vector's nodes in previous subgraph V_s-1 */
    U_s ← CALCULATE (E_s + MAX (V_s-1, C_s-1))
    /* Calculate the Value of each node in subgraph accounting only for
            interconnected nodes and ignoring orphan nodes */
    X_s ← CONCATENATE (E_s, MAX (U_s, X_s-1))
                    /* Final feature vector for the subgraph */
    L_s ← MAX_VALUE-RISK (X_s , L_s-1)

  for a ∈ A do
                  /* Initializing the nodes with feature vector */
                        L_a ← MIN_Prob (X_a , X_a-1)
      while r <= R do
        /*Passing Concatenate all the outputs of imminent adjacent nodes*/
                I_s ← NEIGHBORS(S)
                N_s ← CONCATENATE (L_a, NEIGHBORS (I_s, r))
        /*Passing the node information at each layer through READOUT */
                X ← CALCULATE (E_r + MAX (V_r-1, C_r-1))
          /* Concatenate all the outputs of General Value N layers */
                G_r ← MAX (G_N_r * GENERAL_EXTRACT (X_r)
                  /* Final Expertise likelihood to be optimum */
                L_r ← Prob (G, APPEND (G_r and N_s)
      end
    G ← G + G_r
    L ← L + L_r
  end
  return X and L
end
```

FIGURE 5: ES Security Compliance Expertise Extraction form of Vectors Algorithm.

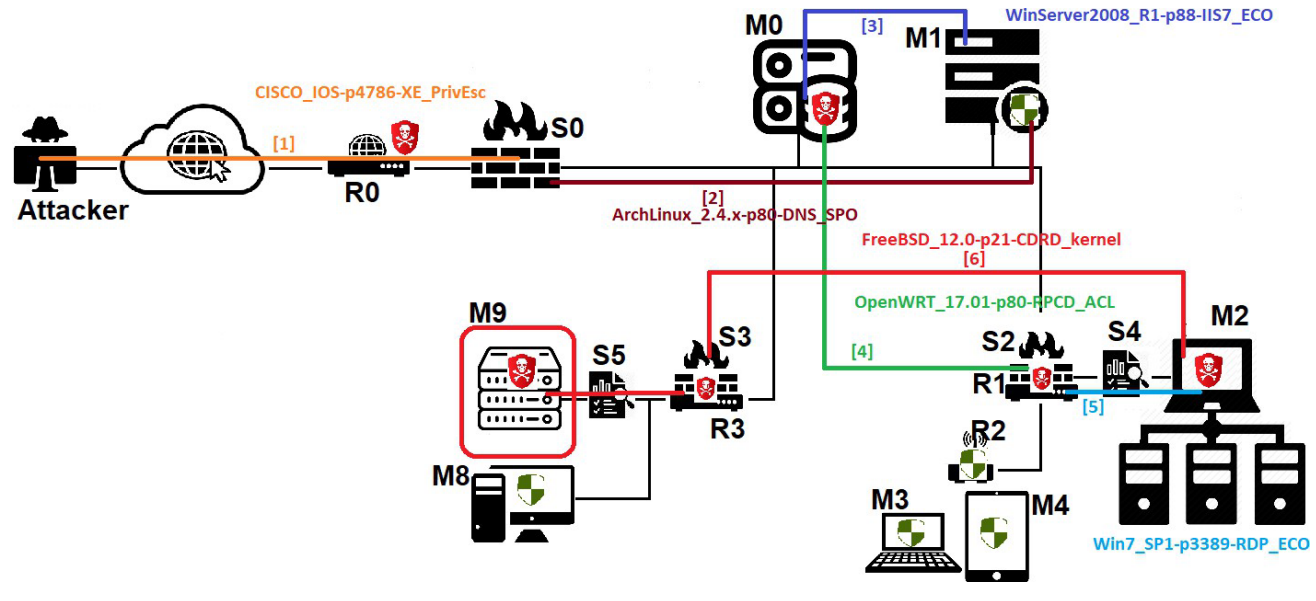

FIGURE 6: Modelling SC activities in the form of attack vectors covering each of the fully assessed machines including reconnaissance, probing, exploiting and privileges escalations

desired, are the data that stimulate execution via the inference engine [27]. In CLIPS ES, rules are defined using the defrule construct and are composed of an antecedent and a consequent. The antecedent of a rule is a set of conditions or conditional elements which must be satisfied for the rule to be applicable [12]. We opted for CLIPS as an efficient approach to implementing our proposed ES as it provides the basic elements of an expert system. The first component of our ES is the domain knowledge composed of fact-list and instance-list which represent the main memory pool for data to be used. The domain knowledge is knowledge about the machine configuration such as operating system, running services, open ports, security defence and storage nature [28].

The second component is the knowledge base which contains all the rules captured, validated and generalized from monitoring human CEH activities and written following the defined rule-base format [19]. The third and last component is the inference engine which is in charge of controlling the overall execution of rules and communicating with the VA or PT tool, respectively, Nessus and Metasploit. The inference engine decides which rules should be executed and then launches the execution. In terms of programming, our ES program written in CLIPS consists of rules, facts, and objects [25] and [17].

Finally, we opted to represent knowledge and expertise directly captured from human CEH in our CLIPS ES through the use of simple or multiple IF-THEN rules. This approach is widely adopted in cyber security in general as it mirrors the real-world situation where the human expert acts (performs a task or sub-task) when a set of conditions are met as illustrated in Figure 6.

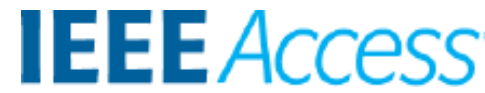

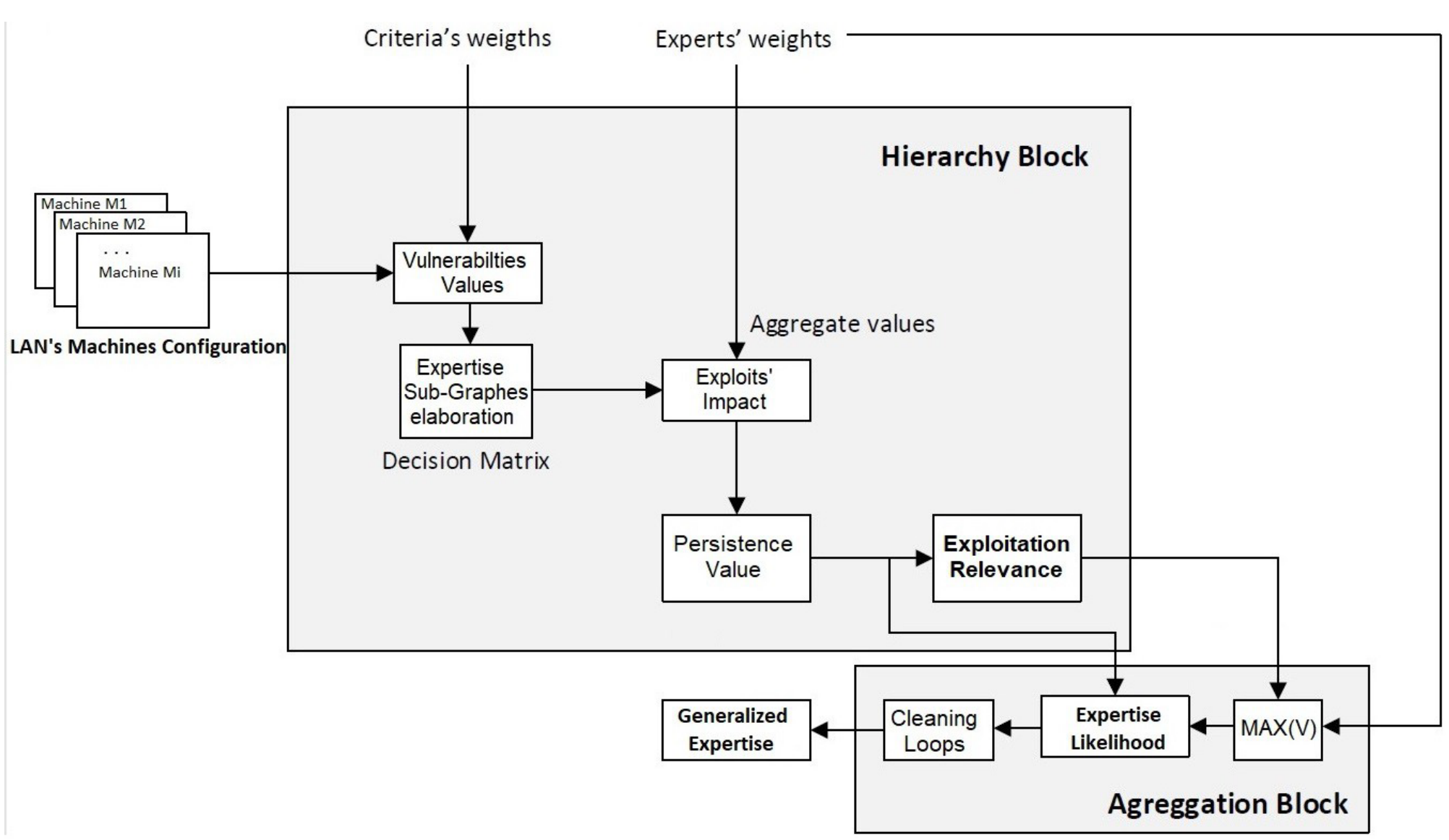

FIGURE 7: Expertise in construction, evaluation and generalization processes.

The first step in implementing the learning process in the form of a decision tree in CLIPS was to decide on which knowledge should be represented and how. Since CLIPS rules' tree should be learned, the tree is also represented as facts and not as rules to make the edition and change in the tree easier [13]. In addition, we opted to use implemented CLIPS rules to traverse the decision tree by implementing the solve tree and learn algorithm following a rule-based approach.

Finally, we utilized the built-in CLIPS pattern-match on facts and objects which can be called from a procedural language, perform its function, and then return control back to the calling program. Therefore, procedural code can be defined as external functions and called from CLIPS. When the external code completes execution, control returns to CLIPS [15], [25].

## IV. PROPOSED ESASCF FRAMEWORK

In this section, we detail the design and implementation of the proposed ESASCF with a special emphasis on the integration of the CLIPS expert system module alongside the processing module. We also discuss virtual test-bed networks' construction out of data collected from real-world corporate networks. This research will produce a proof-of-concept (PoC) framework along with its practical implementation which will assist the human expert in performing security compliance in an efficient and effective manner.

In practice, security compliance activities vary from case to case but generally start with the information-gathering phase, where the expert explores the web using open-source intelligence (OSINT) tools and techniques to gather information about the target system. This later was implemented in an independent data gathering, processing and structuring module during our past research work which we will reuse directly as part of ESASCF [21], [26].

We developed several scripts in C integrated into CLIPS which is a bidirectional Python to C language Foreign Function Interface (CFFI) that facilitates the translation of CLIPS capabilities within the Python ecosystem [24]. These scripts are used to capture certified human experts' (Certified Ethical Hackers or Certified Information Systems Security Professionals) decisions along with the asset parameters that made the human expert make such decisions. For legal and ethical purposes, we also enabled the human expert to assess and control the ESASCF autonomous functioning in order to validate or reject the made decision. Figure 7 shows the proposed rule-based ES functioning in terms of capturing, processing, validating, generalizing and storing expertise for future usage [26].

### A. ESASCF ARCHITECTURE

In ESASCF, we opted for a modular framework that covers the security compliance activities and this is through all VA and PT tasks and sub-tasks. The choice is justified by the nature of VA and PT activities. Figure 5 illustrates the proposed ES-led Automated Security Compliance Framework (ESASCF) including the pre-possessing, rule-based expert system and the VA/PT core. The system consists of the VA module, RBES and memory module as well as the proactive testing and auditing systems module incorporating the interface, Metasploit and Nessus. These modules are represented in Figure 10 [26].

The framework development started by building the first module based on the existing ESASCF which is our previous research work output [21]. The vulnerability assess-

```
(deffacts MAIN::Machine_Status-rules
   (rule (if Machine_Status is ON)
         (then Next_Action is Port_Scan))

   (rule (if Machine_Status is UNKNOWN)
         (then Next_Action is Machine_Status))

   (rule (if Machine_Status is OFF and
             NET_Footprint is TRUE)
         (then Next_Action is Change_Scanning_pivot))

    (rule (if Machine_Status is OFF and
             NET_Footprint is FALSE)
         (then Next_Action is Stop_Scanning))

(deffacts MAIN::Machine_Status
   (goal (determine machine-status)))

(deffacts MAIN::Port_Status-rules
   (rule (if Port_Status is OPEN)
         (then Next_Action is Service_Detect))

   (rule (if Port_Status is FILTERED)
         (then Next_Action is Port_ByPass-Scan))

   (rule (if Port_Status is UNKNOWN)
         (then Next_Action is Port_Re-Scan))

   (rule (if Port_Status is CLOSED and
             Service_Traffic is TRUE)
         (then Next_Action is Change_Probing_pivot))

    (rule (if Port_Status is CLOSED and
             Service_Traffic is FALSE)
         (then Next_Action is Stop_Probing))

(deffacts MAIN::Port_Status
   (goal (determine port-status)))

(deffacts MAIN::Service_Detect-rules
   (rule (if Service_Detect is TRUE)
         (then Next_Action is Vuln_Detect))

   (rule (if Service_Detect is UNKNOWN)
         (then Next_Action is Service_Re-Detect))

   (rule (if Service_Detect is FALSE and
             Port_Status is OPEN)
         (then Next_Action is Change_Detect_pivot))

   (rule (if Service_Detect is FALSE and
             Port_Status is FILTERED)
         (then Next_Action is Change_Detect_Mode))

    (rule (if Service_Detect is OFF and
             Port_Status is CLOSED)
         (then Next_Action is Stop_Detection))

(deffacts MAIN::Service_Detect
   (goal (determine service-detect)))

(deffacts MAIN::Vuln_Detect-rules
   (rule (if Vuln_Detect is TRUE)
         (then Next_Action is Vuln_Exploitation))

    (rule (if Vuln_Detect is UNKNOWN)
         (then Next_Action is Vuln_Re-Detect))

   (rule (if Vuln_Detect is FALSE)
         (then Next_Action is Change_Detect_Script))

   (rule (if Vuln_Detect is FALSE and
             Service_Vuln is TRUE)
         (then Next_Action is Change_Detect_Script))

    (rule (if Vuln_Detect is FALSE and
             Service_Vuln is FALSE)
         (then Next_Action is Stop_Vuln-Assessment))

(deffacts MAIN::Vuln_Detect
   (goal (determine vuln-assessment)))
```

FIGURE 8: An example of expert system rules definition on CLIPS covering PT and VA tasks

ment module uses input data from information gathering, discovery and vulnerability assessment phases to represent it as POMDP (partially observable Markov decision process) environments. The second core component of the framework is the expert system and framework memory. In this module, we opted to represent knowledge in CLIPS through the use of simple or multiple IF-THEN rules which the widely used in expert systems and security programs in general. This approach mirrors the real-world situation where the human expert acts (performs tasks or sub-tasks) when a set of conditions are met [26].

Vulnerability assessment data is collated with all data acquired and formatted during the pre-processing and feature extraction functions which work together as independent scripts. The ES interact directly with ESASCF-memory which serves as the main memory for the framework and the expert system in charge of expertise capturing, generalization, storing and replaying. In ESASCF, Metasploit and Nessus are considered as an entire module of ESASCF and consist of interfaces, libraries, MSF modules, tools and plugins which all will be controlled by the ESASCF through Python scripts relying on CLIPS.

Finally, it is worth mentioning that in our proposed rule-based expert system we opted for using the graphical user interface (GUI). We implemented a simple exchange and display mechanism between the expert system ES, Metasploit MSF and human expert using Python scripts and temporary text files.

## V. TESTING, RESULTS AND DISCUSSION

### A. SETUP OF EXPERIMENTS

The experiments are run on an HP Z2 tower with an Intel CPU Xeon processor E7-4809v3, 8 core, 20MB cache and 2.00GHz, an Un-buffered memory of 64GB DDR4, and graphical NVIDIA P4000 of 8GB. This machine runs Linux Calculate 20 kernel 5.4.6 which is a fast and resource-efficient Linux distribution based on Gentoo and maintains an optimal balance between state-of-the-art processing libraries and renowned stability. The rule-based expert system is developed in CLIPS 6.40 and with the help of CLIPS which is Python CFFI binding that enables us to translate CLIPS capabilities within the Python ecosystem. Furthermore, we implemented all of our memory and data handlers in Python [26].

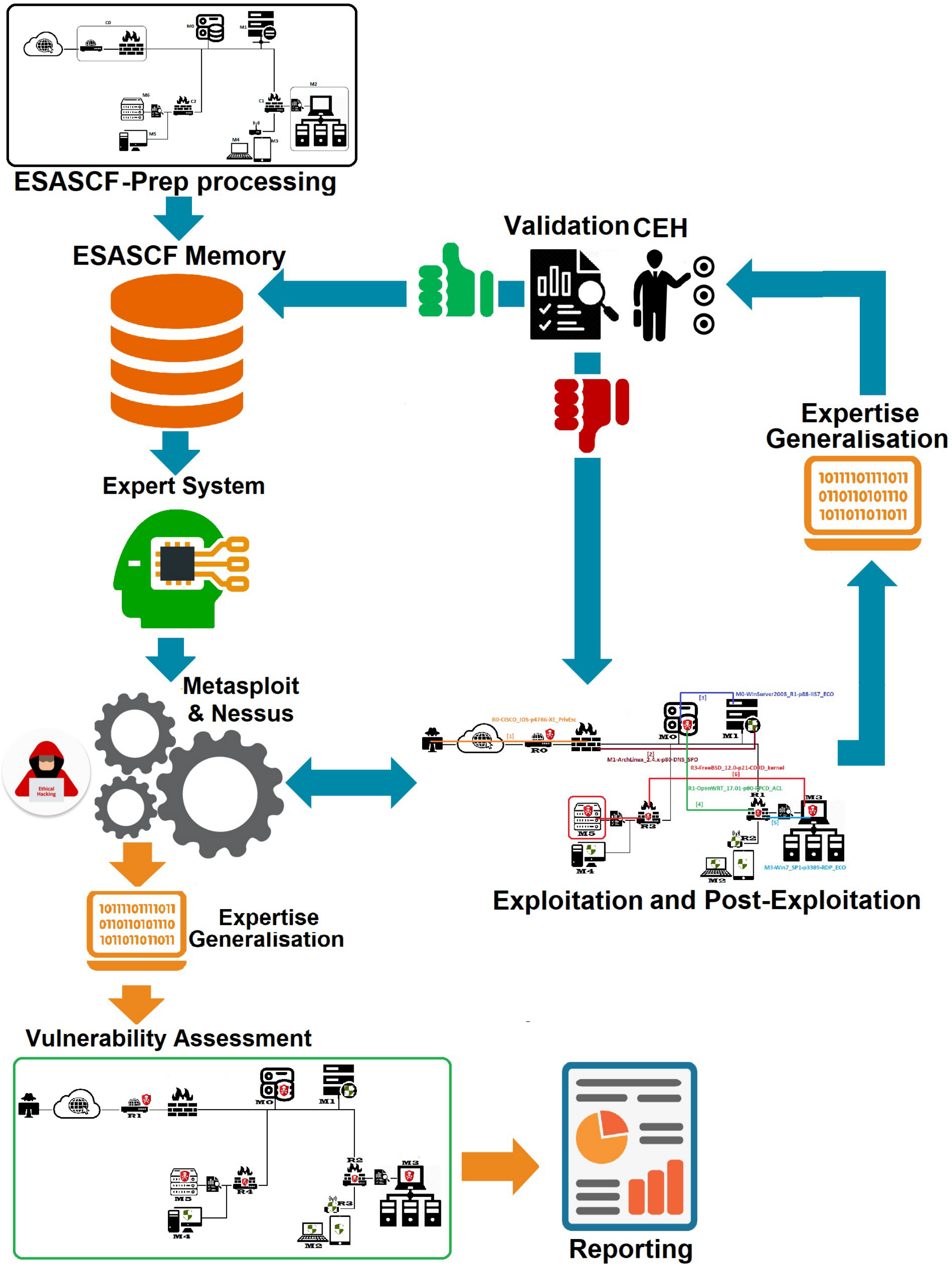

FIGURE 9: ESASCF expertise extraction, validation and generalization workflow for PT and VA

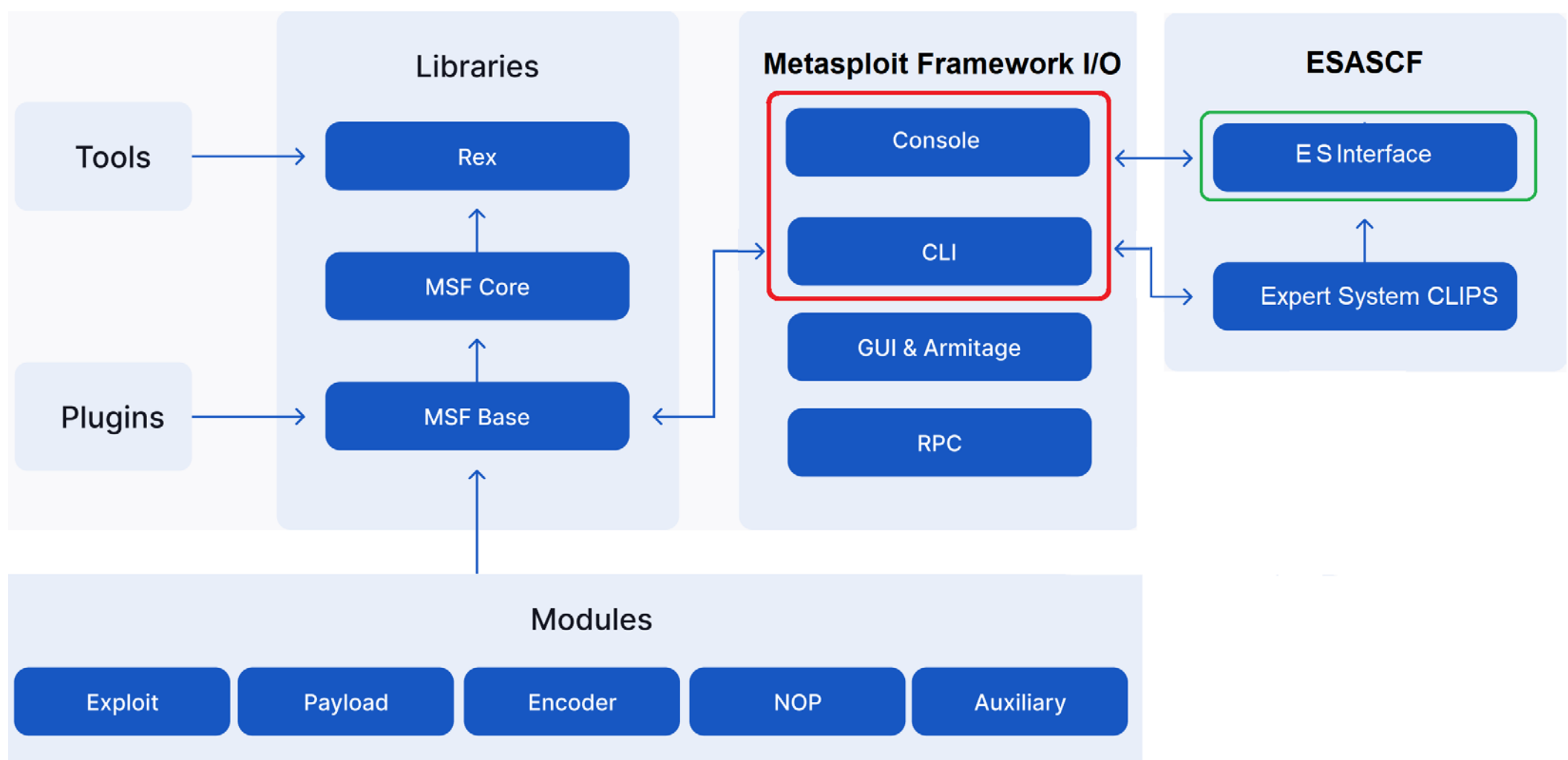

FIGURE 10: Metasploit interaction with ESASCF framework

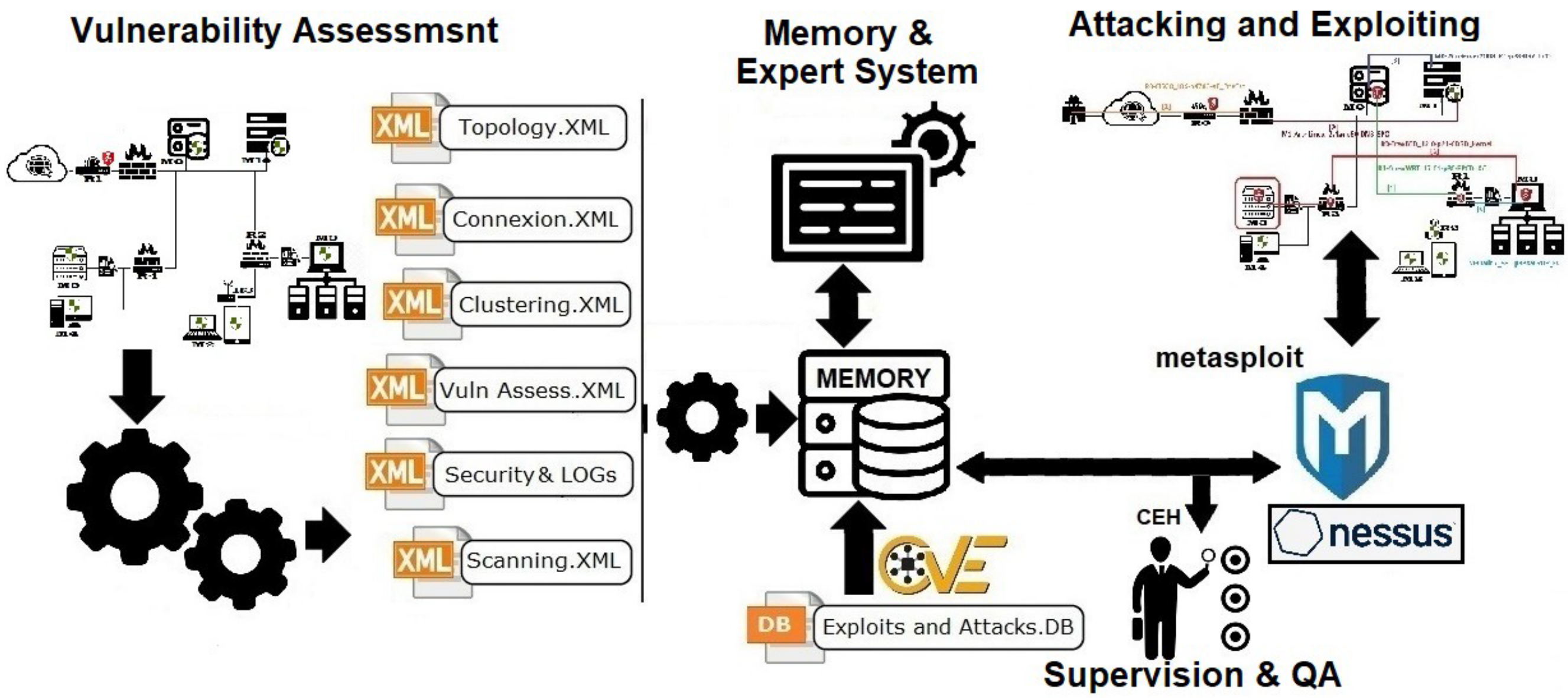

FIGURE 11: Proposed ESASCF framework overall architecture [26].

### *B. RESEARCH DATA INPUT*

This section aims to describe the method used in our research to collect data from real LANs and recreate equivalent virtual networks to be then used to test and validate the ESASCF framework. The starting point which serves as input for this research is 53 different-sized virtual LANs which were recreated out of data imported from real financial institution networks. The collected data include networking, functioning and security data which was used to recreate the virtual equivalent of these networks in a virtual box platform. Computer machines and servers were included in the virtual networks by directly downloading virtual equivalent from a specialized open-source website 'vulnhub.com' which serves as a repository and provides materials that allow ethical hackers to experience digital security, computer software and network administration using virtual appliances.

Security mechanisms including firewalls, Routers and intrusion detection systems were also imported along with the associated configurations (implemented security policy) and included in the virtual networks by adopting a specific approach of considering them as machines and forcing the traffic to transit through them in a specific way to reflect the real-world scenarios. This approach was unavoidable as the virtual environment is restricted in terms of networking.

To sum up, we constructed 53 different networks with size varying from 2 to 250 machines and were categorized as follow: 2-50 small LANs, 55-100 medium LANs and 105-250 large LANs. Even though our research focuses on medium and large networks, we were obliged to start from a small LAN to test the framework. Finally, it is worth mentioning that the 250-machine limitation is purely for operational purposes and larger LANs can be also accommodated with adequate hardware. Figure 12 shows an example large LAN [26].

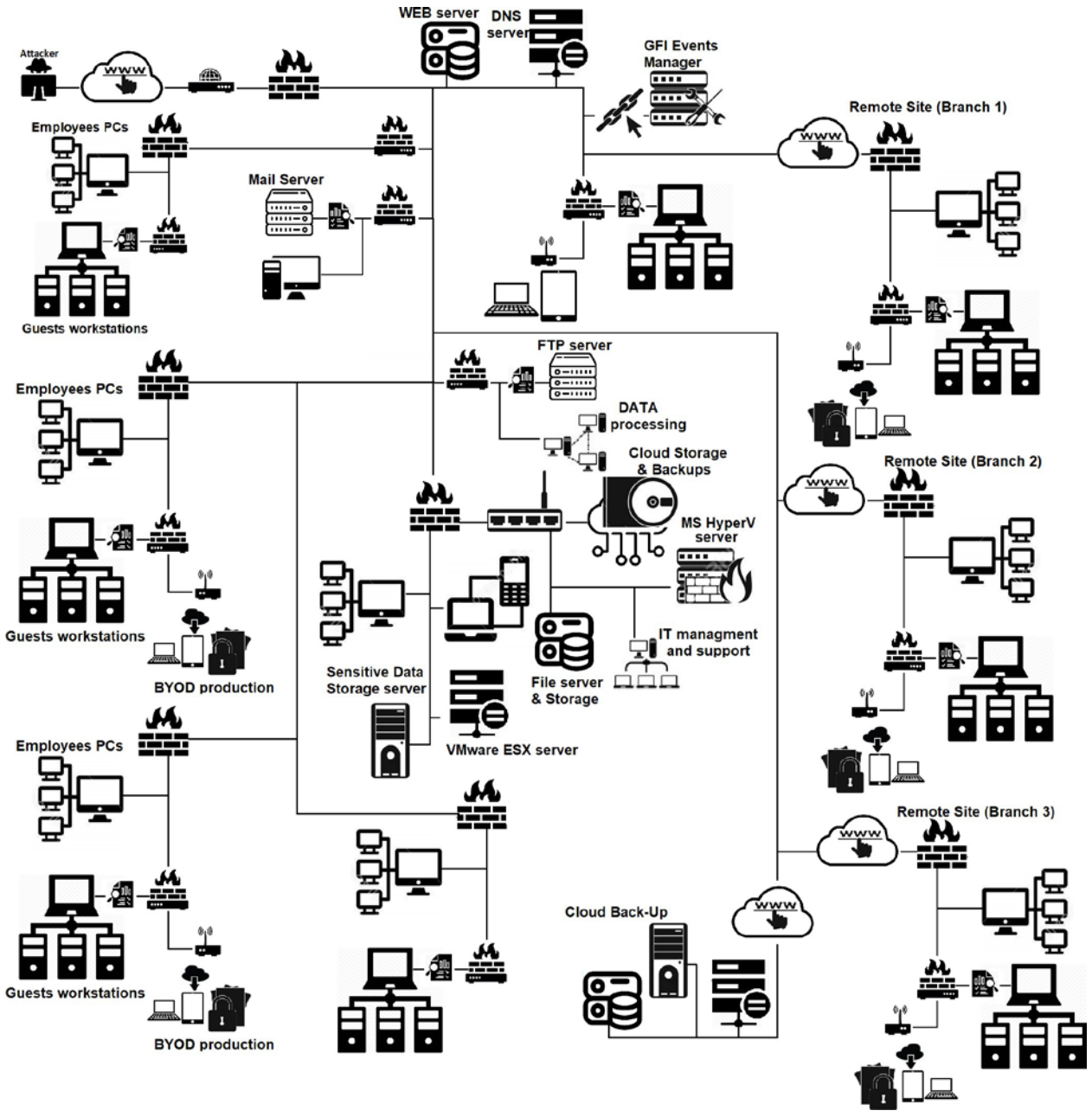

FIGURE 12: Example of a large LAN network used as a test-bed for this research [26]

### C. EVALUATION AND OPTIMIZATION CRITERIA

Currently, security auditing and compliance including PT and VA efficiency is measured following several quantitative and qualitative metrics which are widely adopted and standardized as performance measurement criteria. Nonetheless, the operational cost and the reliability of the results remain the most relevant ones. In terms of relevance and accuracy, we elaborated a hierarchical function that calculated the value of expertise extracted and its relevance alongside the extraction process outlined in Figure 8. To tack we assume that security testing and auditing tools and system licensing constitute 1/10 of the total cost [10]. The remaining cost is allocated to pay human experts conducting compliance assessing and testing activities [2], [23]. Therefore we simplified the efficiency evaluation metric to only account for the average running time (which is reflected in cost as experts are often hourly paid). The second metric is compliance coverage measured by the number of performed assessment and tests which are measured here by the number of covered machines including low-risk machines often neglected by human experts and which ESASCF cover fully.

## VI. EXPERIMENTAL RESULTS AND DISCUSSION

ESASCF testing was carried out in two stages. First, we tested the framework efficiency in different security compliance situations when ESASCF observed and captured expertise from human CEH performing initial VA and PT using Nessus and Metasploit respectively, and then ESASCF was used to repeat the security compliance after a few changes were introduced.

### A. OBTAINED RESULTS

Figures 13 and 14 illustrate the huge contribution of ESASCF in compliance scenarios when VA and PT are repeated periodically or after introducing a few changes (e.g. 25%). The impact in terms of time is less significant in VA as the assessment practice is more deterministic and more automated. Nonetheless, the re-testing efficiency enhancement is far more important with the practice running time representing, in large LANs, a fifth (1/5) of the normal time required for testing when only 25% or less of configuration change has been introduced to the LANs which in fact represent the real-world situation and more-likely situation in IT.

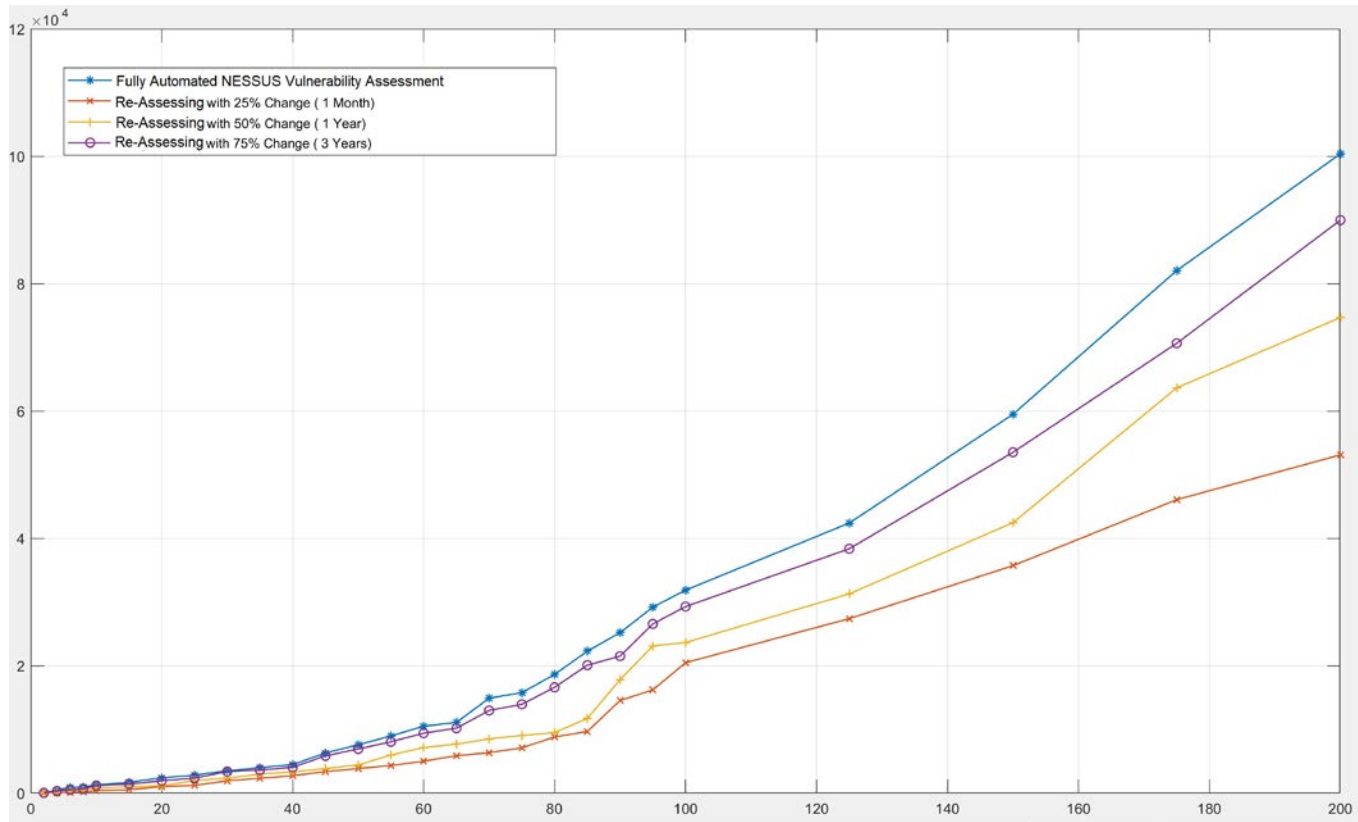

FIGURE 13: ESASCF performances in network vulnerability re-assessing using Nessus on different size LANs

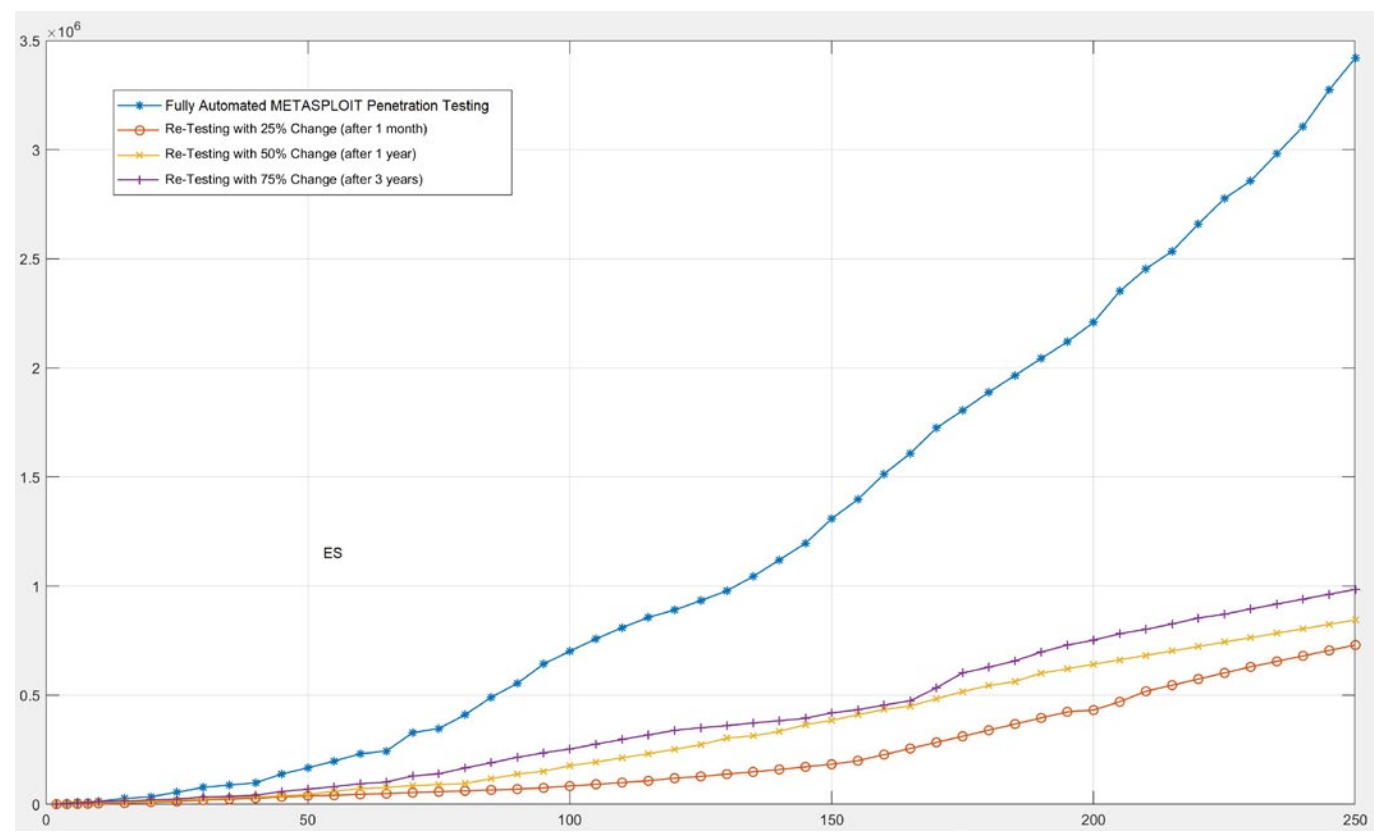

FIGURE 14: ESASCF performances in network Penetration Re-Testing using Metasploit on different size LANs

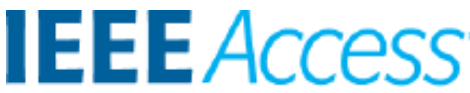

Finally, we compared the ESASCF performances with full blind automation and human expert CEH performances in terms of retesting the same LANs after introducing the time 25% changes. Figure 15 illustrates the obtained results.

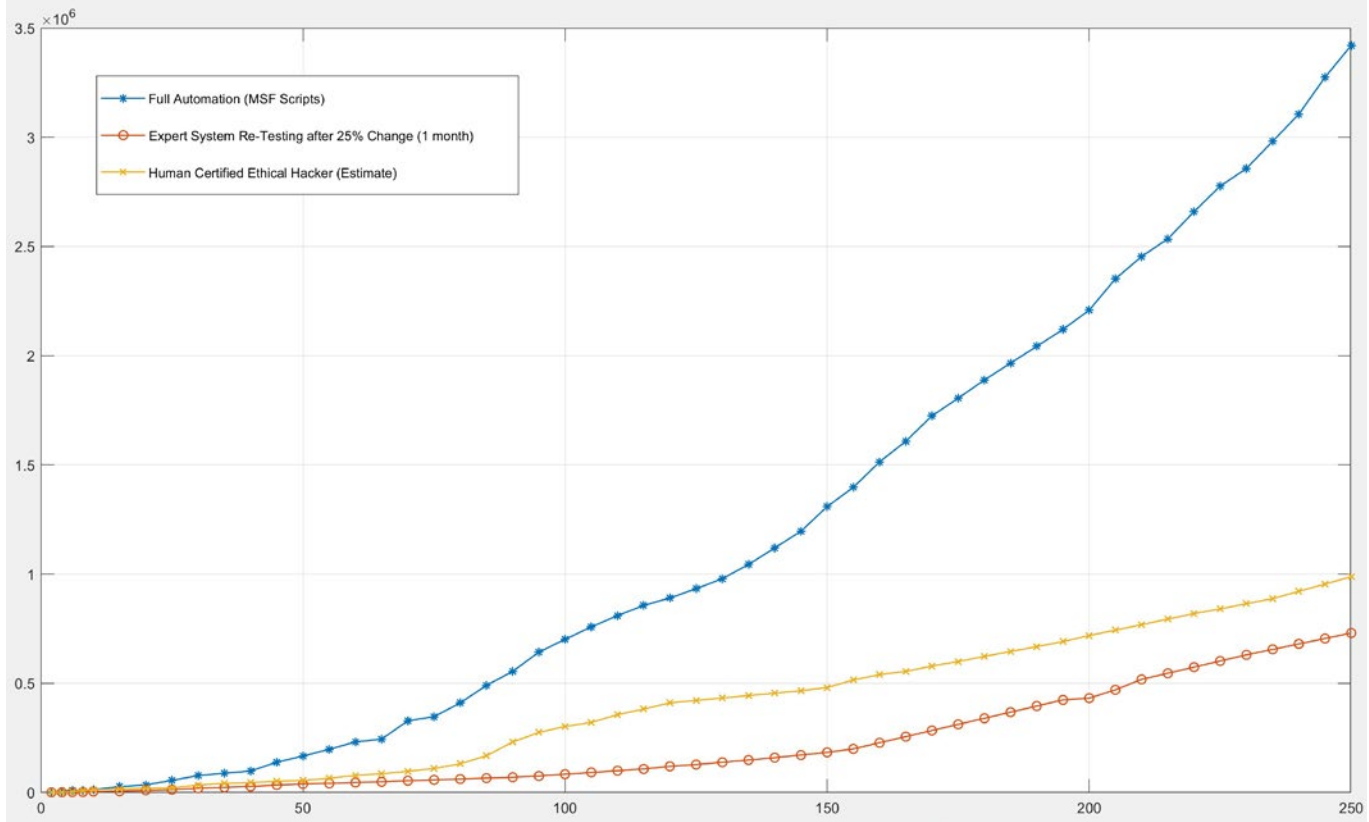

FIGURE 15: ESASCF performance comparison with blind automation PT and human expert (CEH) PT in different network sizes

From the results, we confirmed that ESASCF outperforms the human expert as well the blind automation which validates the contribution of ES-led security compliance. The unanimous results reflect the contribution of expertise capturing and reuse in cyber security compliance. In addition to the quantitative results brought by ESASCF to the security compliance practice and specifically vulnerability assessment (VA) and penetration testing (PT), the proposed ES-led solution produces a similar compliance quality as with highly qualified and certified human experts. Figure 16 illustrates the qualitative impact of ESASCF on the security compliance practice notably by enabling high-quality expertise extraction and reuse. The results clearly show that ESASCF security testing coverage outperforms any human expert along with attack coverage far larger and more precise in the sense that only the relevant scenarios were covered which in the large network includes running 15 exploits, 6 post-exploitation payloads and resulted in compromising five high-value targets computer or servers as illustrated in Figure 16 where each coloured line represent an extracted and validated attack vector [45].

## VII. CONCLUSION AND FUTURE WORKS

This paper investigated the enhancement of security compliance performances through the use of a rule-based expert system within the industrial VA and PT tools and systems. This enables industrial systems to acquire, generalize and re-use the expertise learned from human experts and prioritize its use in future relevant scenarios notably similar cases and re-testing/ re-assessing. The proposed ESASCF is based on an expertise identification and extraction model and covers all networks and infrastructures VA and PT which optimize the SC practice and enhance the efficiency and effectiveness of current industry tools and systems such as Metasploit and Nessus.

The main contribution of the proposed framework built upon the introduced model is to safely replace (or minimize) the human expert intervention in the SC practice and make it accessible to non-experts. On the other hand, ESASCF allows efficient and accurate SC in terms of consumed time, testing coverage, resource use and impact on the assessed assets. The obtained results show that ESASCF defeats human-led and fully automated security compliance assessing and testing performances in terms of consumed time which reflects the cost of the practice in general. This improvement is particularly obvious in the medium and large network contexts.

The learning process is the second strength of the proposed model notably in the case of re-assessing and retesting the same LAN after a few changes were introduced which represent the real-world context in security. Here again, the performance enhancement and the previously extracted expertise reuse are enormous, especially in large LANs which is translated into further performance and practically confirms the suitability of our proposed approach.

Finally, despite the fact that this work opened the door for the use of ES-led security compliance, the proposed framework can be further enhanced notably by addressing current limitations of CLIPS namely the single-level rule sets which pushed us to arrange rule sets in a hierarchy for loop sub-task such as the port probing and service detection. The second issue faced in CLIPS is the issue related to matching rules and objects as it is not possible to embed rules in objects which remain problematic in some aspect of security compliance such as changing pivot for re-scanning or re-testing. In addition, the CLIPS lacks an explicit agenda mechanism making forward chaining the only available approach to control flow and therefore pushing toward manipulating tokens in working memory as the only alternative to implementing other kinds of reasoning. One of the future improvements is the migration of the ES towards NExpert Object which is highly reliable and portable. It also includes facilities for designing graphical interfaces and enables the use of script language in the front end. Furthermore, AI presents some limitations in automating VA and PT notably the heavy dependence on human expertise and knowledge in the processing and validation which in practice limit the amount of automation and many researchers are directing the use of additional ML techniques such as Reinforcement Learning to achieve such full automation.

## FOOTNOTES

### ETHICAL APPROVAL

This research was deemed as not requiring the University's Ethical Committee Approval as it doesn't fall under any of the cases requiring ethical approval.

### FUNDING

No Funding.

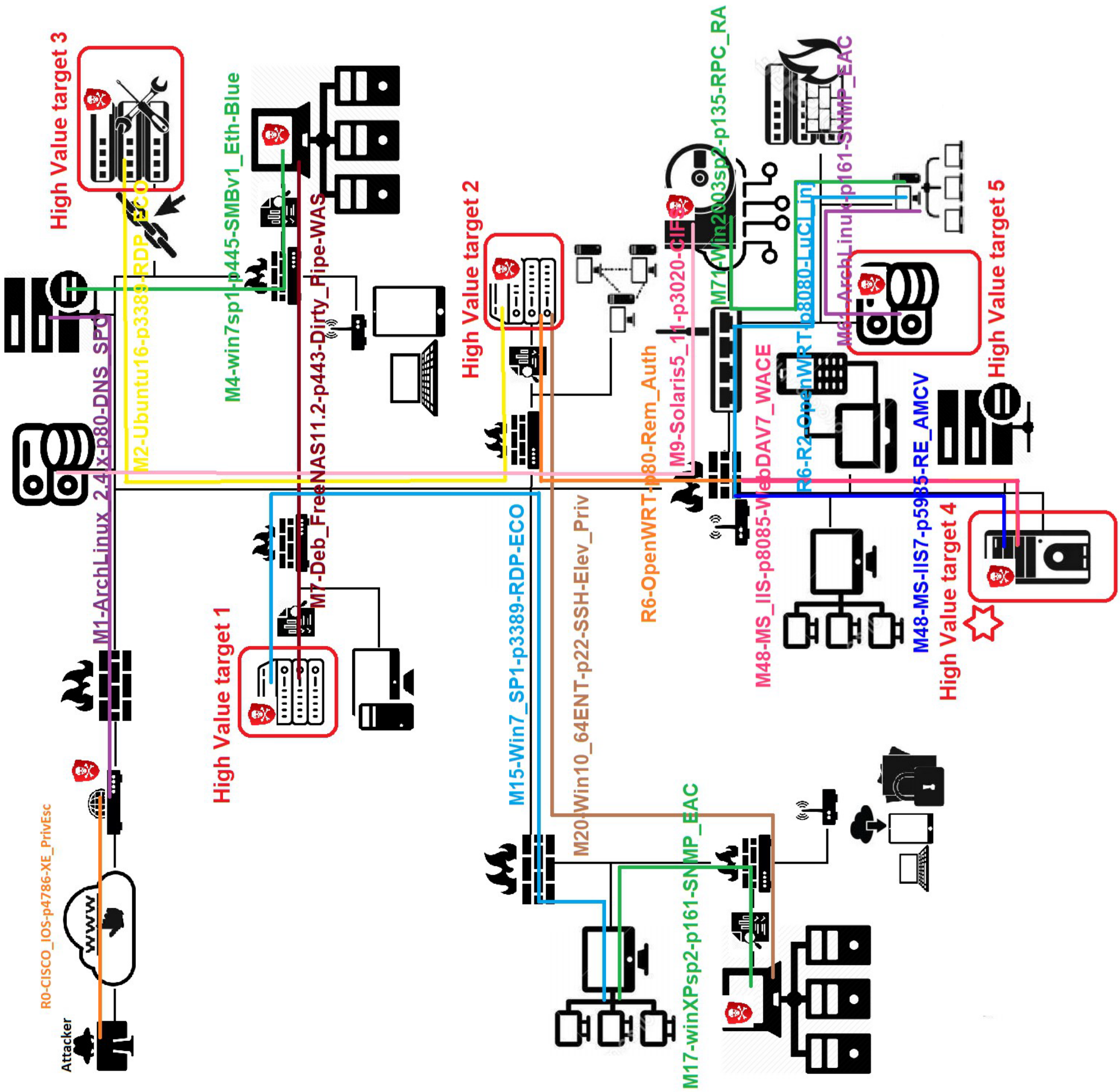

FIGURE 16: An example of ESASCF expertise extraction, generalization in a large LAN context

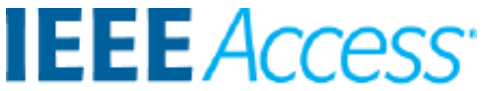

### AVAILABILITY OF DATA AND MATERIALS

ESASCF code and Virtual LANs data sets generated for this research are available upon request.

### COMPETING INTERESTS

The authors declare that they have no known competing interests or personal relationships that could have appeared to influence the work reported in this paper.

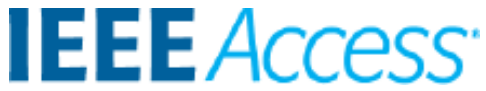

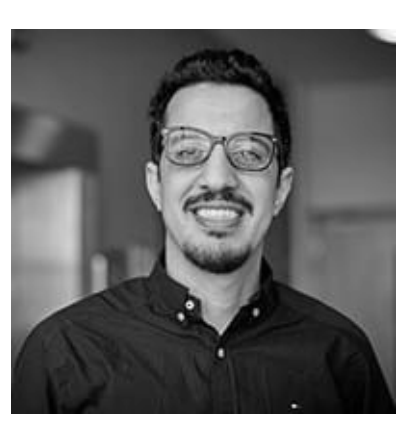

Dr Mohamed Chahine Ghanem is an Associate Professor and deputy director of the Cyber Security Research Centre, he holds an MSc in Digital Forensics with Distinction and a PhD in Cyber Security from the City, University of London. He is a Senior Fellow of HEA and holds a PGDip in Security Studies. Mohamed-Chahine is an IEEE member and a professional member of the British Computer Society achieved many certificates, such as GCFE, CISSP, ACE, XRY and CPCI with over 15 years of experience in the field of cybersecurity, digital forensics and incident investigation at the law-enforcement and corporate level. His research focuses on applying AI to solve real-world cybersecurity and digital forensics problems and published numerous research papers in the world's top cybersecurity journals.

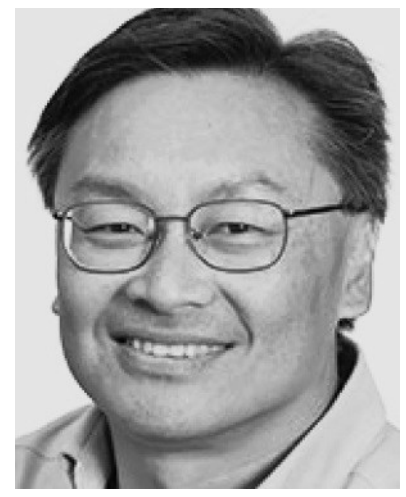

Professor Thomas M. Chen is currently a Professor in cyber security at the City, University of London. He was previously a Senior Technical Member of Staff with the GTE Laboratories (now Verizon), Waltham, Massachusetts. He joined Southern Methodist University, Dallas, as an Associate Professor, and then Swansea University, Wales, as a Professor in networks. He has served as the Former Editor-in-Chief for IEEE Communications Magazine, IEEE Network, and IEEE COMMUNICATIONS SURVEYS AND TUTORIALS. He was a co-recipient of the Fred W. Ellersick Best Paper Award, in 1996. He has written or edited seven books, and 41 book chapters, and owns two U.S. patents. His research interests include cyber security, online extremism, and computer networks

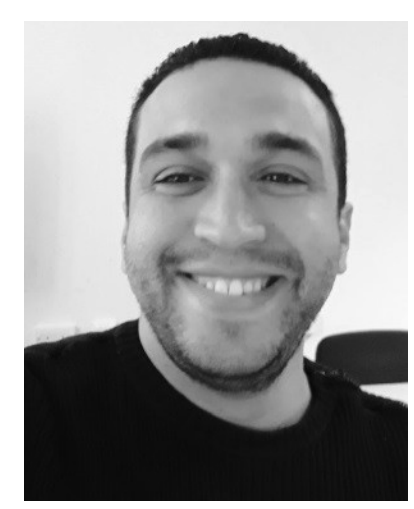

Dr. Mohamed Amine Ferrag earned his Bachelor's, Master's, Ph.D., and Habilitation degrees in Computer Science from Badji Mokhtar—Annaba University in Annaba, Algeria, completing his studies in 2008, 2010, 2014, and 2019, respectively. He served as an Associate Professor in the Department of Computer Science at Guelma University, Algeria, from 2014 until 2022. Concurrently, from 2019 to 2022, he held the position of Senior Researcher at the NAU-Lincoln Joint Research Center of Intelligent Engineering, based at Nanjing Agricultural University in China. As of 2022, Dr. Ferrag is the Lead Researcher at the Artificial Intelligence & Digital Science Research Center at the Technology Innovation Institute in Abu Dhabi, United Arab Emirates. Dr Ferrag's research is primarily focused on a spectrum of topics within the cyber security domain, including wireless network security, network coding security, applied cryptography, blockchain technology, generative AI, software security, and the application of AI in cyber security. His scholarly output includes over 140 papers published in international journals and conference proceedings. Dr. Ferrag has spearheaded numerous projects in research and development, fostering collaborative ties with academic institutions in the UK, Australia, and China. His contributions to the field include the creation of two cybersecurity datasets, which have become essential resources for AI researchers worldwide. His academic contributions have been recognized with the 2021 IEEE TEM Best Paper Award, the 2022 Scopus Algeria Award, and many best paper conference awards. He has consistently been named on Stanford University's list of the world's top 2% of scientists four times from 2020 through 2023. Dr. Ferrag also contributes to the academic community as an associate editor for prestigious journals, such as the IEEE Internet of Things Journal and ICT Express (Elsevier). In addition to his research and editorial roles, Dr. Ferrag is a Senior Member of the Institute of Electrical and Electronics Engineers (IEEE).

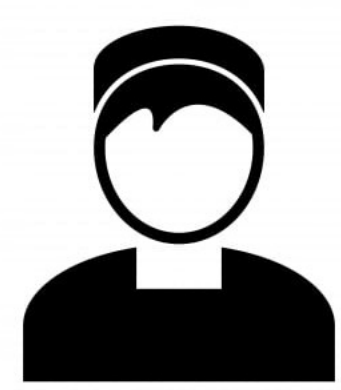

Mr Mohyi-Eddine Kettoche is a visiting researcher at the Cyber Security Research Centre at London Metropolitan University. Mr Kettouche Holds an Engineering degree in Computer Sciences and a Master's Degree in Applied Computing from Claude Bernard University Lyon 1 and has more than 10 years of experience in developing innovative solutions for businesses and organisations. His research primarily focuses on Applied AI in network security, information retrieval and extraction in big data, expert systems, and blockchain technology.